\documentclass[a4paper,fleqn,usenatbib]{mnras}
\usepackage{newtxtext,newtxmath}

\usepackage[T1]{fontenc}
\usepackage{ae,aecompl}
\usepackage{enumerate}
\usepackage{soul} 
\usepackage{lineno}

\usepackage{comment}
\usepackage{graphicx}	
\usepackage{amsmath}	
\usepackage{amssymb}	
\usepackage{ulem}
\usepackage{color}
\newcommand{\add}[1]{\textcolor{black}{#1}}

\providecommand{\nat}[0]{Nature}

\providecommand{\apj}[0]{Astrophys. J.}
\providecommand{\apjl}[0]{Astrophys. J. Lett.}
\providecommand{\apjs}[0]{Astrophys. J. Supp. Ser. }

\providecommand{\aap}[0]{Astron. Astrophys. }

\providecommand{\araa}[0]{Ann.\ Rev. Astron. Astroph. }
\providecommand{\physrep}[0]{Phys. Rep. }
\providecommand{\mnras}[0]{Mon. Not. Roy. Astron. Soc. }

\providecommand{\prd}{Phys. Rev. D.}

\providecommand{\physrep}[0]{Phys. Rep.}

\title[GW+CTA]{Strategies for the Follow-up of Gravitational Wave Transients with the Cherenkov Telescope Array}

\author[I. Bartos et al.]{I. Bartos,$^{1,2}$\thanks{E-mail: imrebartos@ufl.edu}
T. Di Girolamo,$^{2,3}$\thanks{E-mail: tristano.digirolamo@na.infn.it}
J.R. Gair,$^4$
M. Hendry,$^5$
I.S. Heng,$^5$
\newauthor
T.B. Humensky,$^2$
S. M\'arka,$^2$
Z. M\'arka,$^2$
C. Messenger,$^5$
R. Mukherjee,$^6$
\newauthor
D. Nieto,$^7$
P. O'Brien$^8$
and M. Santander$^{6,9}$
\\
$^1$Department of Physics, University of Florida, Gainesville, FL 32611, USA \\
$^2$Department of Physics, Columbia University, New York, NY 10027, USA \\
$^3$Dipartimento di Fisica ``Ettore Pancini" dell'Universit\`a ``Federico II" and Istituto Nazionale di Fisica Nucleare, 80126, Napoli, Italy \\
$^4$School of Mathematics, University of Edinburgh, Edinburgh EH9 3FD, United Kingdom \\
$^5$SUPA, University of Glasgow, Glasgow G12 8QQ, United Kingdom \\
$^6$Department of Physics \& Astronomy, Barnard College, Columbia University, New York, NY 10027, USA \\
$^7$Facultad de Ciencias F\'isicas, Universidad Complutense de Madrid, 28040 Madrid, Spain\\
$^8$Space Research Centre, Department of Physics and Astronomy, University of Leicester, Leicester LE1 7RH, United Kingdom \\
$^{9}$Department of Physics \& Astronomy, University of Alabama, Tuscaloosa, AL 35487, USA
}

\begin{document}
\label{firstpage}
\pagerange{\pageref{firstpage}--\pageref{lastpage}}
\maketitle
%
\begin{abstract}
The observation of the electromagnetic counterpart of gravitational-wave (GW) transient GW170817 demonstrated the potential in extracting astrophysical information from multimessenger discoveries. The forthcoming deployment of the first telescopes of the Cherenkov Telescope Array (CTA) observatory will coincide with Advanced LIGO/Virgo's next observing run, O3, enabling the monitoring of gamma-ray emission at E $>\,$20 GeV, and thus particle acceleration, from GW sources. CTA will not be greatly limited by the precision of GW localization as it will be be capable of rapidly covering the GW error region with sufficient
sensitivity. We examine the current status of GW searches and their follow-up effort, as well as the status of CTA, in order to identify some of the general strategies that will enhance CTA's contribution to multimessenger discoveries.
\end{abstract}



\section{Introduction}

The Advanced LIGO observatories' first two observing \add{runs} saw an extensive effort to search for multimessenger emission from gravitational wave (GW) sources, covering both the electromagnetic and neutrino spectra \citep{2016ApJ...826L..13A,2016PhRvD..93l2010A,2017PhRvD..96b2005A,2017ApJ...848L..12A,2017ApJ...850L..35A}. This effort culminated, on 2017 August 17, with the observation across the electromagnetic spectrum of a gamma-ray burst (GRB) and a kilonova as a consequence of a binary neutron star merger which was detected in GWs \citep{2017PhRvL.119p1101A, 2017ApJ...848L..12A,2017ApJ...848L..13A}. The Advanced Virgo interferometer \citep{2015CQGra..32b4001A} was also operational at this time and aided the discovery, reducing the sky localization region. The success of the observational campaign for this event, which marked the start of multimessenger astronomy with GWs, shows the importance of coordinated follow-up observations and of the strategies to carry out them. In the next few years, the LIGO and Virgo detectors will continuously improve their sensitivity \citep{2016LRR....19....1A}, while additional GW interferometers are envisaged to come online \citep{PhysRevD.88.043007,LIGOIndia}, promising the regular detection of a variety of sources with potential multimessenger signatures.

The Cherenkov Telescope Array (CTA, \citealt{2013APh....43....3A}) will soon expand the multimessenger observational horizon with an unprecedented sensitivity to sources producing very high-energy ($>$\,10 GeV) gamma-ray emission \citep{1998PhR...305...93O}. The first CTA telescopes are envisaged to come online in 2019, bringing the first joint CTA-GW searches during the LIGO/Virgo third observing period. CTA will continuously increase its sensitivity as more telescopes are installed, broadening its reach in parallel with increased GW capabilities.

Multimessenger sources of interest include the formation of black hole--accretion disk systems that drive relativistic outflows, giving rise to high-energy emission. Such a system can arise from the formation or merger of compact objects, such as neutron star--neutron star or black hole--neutron star mergers \citep{2010CQGra..27q3001A}, core collapse supernovae with rapidly rotating cores \citep{2013CQGra..30l3001B}, and plausibly from binary black hole mergers \citep{2016arXiv160204226S,2016ApJ...819L..21L,2016ApJ...822L...9M,2016ApJ...821L..18P,2017arXiv170102328B,2017ApJ...835..165B}. The resulting black hole--accretion disk system then drives a relativistic outflow, and dissipation within the outflow can accelerate cosmic rays and produce non-thermal, high-energy gamma-ray \citep{2004RvMP...76.1143P,2012Sci...337..932G} and neutrino emission \citep{1997PhRvL..78.2292W,2013RvMP...85.1401A}.

It is currently unclear how high in energy gamma-ray emission can reach from multimessenger sources of interest such as GRBs. The Large Area Telescope on the {\sl Fermi} satellite ({\sl Fermi}-LAT) has detected GRB gamma rays up to tens of GeV energies \citep{2013ApJS..209...11A}, with no clear cutoff, and with the limitation that the universe becomes opaque at the highest energies for sources at typical GRB distances. Ground-based, imaging atmospheric Cherenkov telescopes (IACTs) have so far made no detection from GRBs \citep{2007ApJ...667..358A,2011ApJ...743...62A}, but this is consistent with the extrapolated high-energy flux from {\sl Fermi}-LAT observations. Nonetheless, the observed ultra-high energy cosmic-ray flux \citep{2011RvMP...83..907L} and cosmic neutrinos \citep{2013Sci...342E...1I,2015arXiv150900983B} show that particle acceleration and high-energy emission reaches much higher energies than the current observational limit.

Joint LIGO/Virgo+CTA observations represent a promising probe to very high-energy gamma-ray emission from extreme cosmic transients. GW detections can unambiguously identify nearby black-hole formation or evolution, and allow CTA to carry out searches that can connect very high-energy emission to the progenitor. While typical transient observations, such as those of GRBs, are at cosmological distances that hinder the detection prospects of very high-energy photons due to photon-photon absorption induced by interactions with the extragalactic background light (EBL), observed GW sources will mostly be within the distance range ($\lesssim 1$\,Gpc) at which the highest energy photons can reach the Earth. While very high-energy emission from sources of interest is uncertain, extrapolating observed GRB emission to higher energies indicates that CTA could easily detect energetic photons from GW sources \citep{2014MNRAS.443..738B}. Furthermore, CTA is well suited to carry out follow-up observations of GW triggers due to its fast response, large field of view and unprecedented sensitivity, enhancing the utility of joint LIGO/Virgo+CTA observation campaigns.

This paper has two objectives. (1) It aims to summarize the status, operation and prospects of GW detectors for the CTA community, and vice versa, in order to provide a concise review of the opportunities and constraints of GW and very high-energy observations. (2) With the near-future onset of joint observations, the paper aims to outline the steps ahead needed to carry out effective multimessenger surveys, and to give specific recommendations that can help optimize this joint effort.

The paper is organized as follows. Section \ref{sec:jointsources} briefly describes the joint sources of interest and the emission mechanisms. Sections \ref{sec:GW} and \ref{sec:CTA} outline the detectors, observation strategies, and plans for GW facilities and CTA, respectively. Section \ref{ref:jointsearch} describes the multimessenger search strategies and prospects. Section \ref{sec:conclusion} presents a summary and lists our recommendations for joint observation campaigns.

\section{Joint Sources of Gravitational wave and High-energy Gamma-ray Emission}
\label{sec:jointsources}

A binary system with total mass up to a few hundred solar masses will generate GWs detectable by Advanced LIGO/Virgo, if the event occurs within the detector's horizon, during the final stages of binary inspiral and merger~\citep{Sathyaprakash2009}. The requirement that the binary system is about to merge ensures that only binary systems containing neutron stars or black holes are potential sources of GW emission for Advanced LIGO. During this final phase leading through merger and into ringdown, the evolution of the binary system is expected to be gravitationally dominated and the GW emission can therefore be accurately predicted by solving the Einstein equations of general relativity. The inspiral can be characterized using post-Newtonian theory, which constructs GW emission by evaluating an expansion of the field equations in powers of the velocity~\citep{2014LRR....17....2B}. The post-Newtonian expansion cannot be used to describe the last few cycles of inspiral and the subsequent merger, where the velocities of the binary components approach a significant fraction of the speed of light. Modeling this portion of the signal requires a full numerical solution of Einstein's field equations on a computer using numerical relativity~\citep{2005PhRvL..95l1101P}. After merger, the two binary components form a single, highly perturbed, black hole, which then settles down to a stationary state through emission of GW radiation as a superposition of damped sinusoids, which is known as the ringdown. Two hybrid waveform families also now exist that smoothly combine the three phases into a single model that is tuned to match numerical relativity simulations~\citep{2016PhRvD..93d4007K,2017PhRvD..95d4028B} and is better suited for use in GW data analysis. The fact that all three phases of the GW signal can be well modeled allows binary systems to be identified in the LIGO/Virgo data set using matched filtering, which significantly improves the distance to which such systems can be observed.

Binary neutron star mergers sweep through the whole of the LIGO/Virgo observation band of $\sim 10$--$1000$\,Hz. The final stages of merger occur at frequencies away from the most sensitive part of the LIGO noise curve and the majority of the signal-to-noise ratio and information comes from the inspiral portion of the signal. 

Binary black hole (BBH) systems have higher masses and hence reach merger at lower frequencies, which for systems with total mass of a few tens of solar masses can be in the most sensitive portion of the LIGO noise spectrum. The majority of the signal-to-noise for higher-mass BBH systems comes from the final stages of inspiral, the merger and subsequent ringdown. The higher mass typical of BBH systems means they can be observed to greater distances, with systems like GW150914 (component masses of $\sim30\ M_\odot$ and $\sim35\ M_\odot$) being detectable to distances of $\sim9$ Gpc by Advanced LIGO at its design configuration~\citep{2016ApJ...818L..22A}. Systems with masses much higher than GW150914 are still potentially detectable by Advanced LIGO, but only during the merger and ringdown phases. These systems will therefore tend to be less well characterized than lighter BBHs~\citep{2017arXiv170404628T}. For systems with total mass above $\sim500\ M_\odot$ only the ringdown signal is in the LIGO frequency band and LIGO's sensitivity to such systems is significantly poorer~\citep{2014PhRvD..89j2006A}.

Core-collapse supernovae (CCSNe) are also potential sources of GW emission~\citep{2016PhRvD..94j2001A}. Significant GW emission occurs only if there is substantial asymmetric acceleration of the stellar material during the supernova. A number of mechanisms generating the required asymmetries have been discussed. The most extensively studied is the presence of significant rotation during core collapse. Rotation generates an axisymmetric oblate deformation of the collapsing core. The extreme acceleration of the material at core-bounce generates a burst of GWs that is linearly polarized. In slowly rotating stars, the rotating core collapse model does not apply, but significant GW emission can be generated by neutrino-driven convection or the standing accretion shock instability. Simulations of CCSNe in 2D and 3D have been used to demonstrate the generation of GWs, but in general the core collapse mechanism is complex such that the GW emission cannot be sufficiently well modeled to generate templates for data analysis. However, the GW emission from a CCSN is expected to be short in duration and broad in spectrum, making these good candidates for LIGO/Virgo burst detection algorithms~\citep{2016PhRvD..94j2001A}.

Short GRBs are thought to be powered by neutron star-neutron star (NS-NS) or neutron star-black hole (NS-BH) mergers (e.g., see~\cite{2014ARA&A..52...43B}). The unambiguous association of GRB170817A with GW170817 recently confirmed this hypothesis at least for some short GRBs, however with a \add{soft} prompt emission extending only to $\sim$1\,MeV~\citep{2017ApJ...848L..13A, 2017ApJ...848L..14G, 2017ApJ...848L..15S}, while at very high energies the H.E.S.S. IACTs set some upper limits at later times~\citep{2017arXiv171005862H}. {\sl Fermi}-LAT has detected emission above $100\ \textrm{MeV}$ from several short GRBs, most notably GRB090510~\citep{2009Natur.462..331A,2010ApJ...716.1178A}. The GeV component is likely produced by inverse Compton scattering, though the nature of the seed photon population is still unsettled and may depend on the environment the GRB is expanding into; possibilities include the synchrotron photons produced by electrons accelerated at the external shock generated when the relativistic jet is decelerated by the external medium~\citep{1994MNRAS.269L..41M,1994ApJ...432..181M}, or prompt radiation emitted at
smaller radii. In either case, the thermal plasma behind the external shock provides the energetic leptons~\citep{2014ApJ...788...36B}. Because {\sl Fermi}-LAT is fluence limited, it is clear that CTA will have the raw sensitivity required to detect a significant population of short GRBs. A time delay may  be possible between a GW trigger and any short GRB emission in the case of a NS-NS merger, if the merger yields a supramassive NS (though \cite{2015PhRvL.115q1101M} argues against the viability of this scenario); in this case the GRB may be delayed by O(10$^3$\,s) with respect to the GW trigger~\citep{1998ApJ...507L..45V,2015arXiv150501420C}, which will need to be taken into consideration in designing the electromagnetic follow-up strategy. While photon-photon absorption can suppress the emission of gamma rays above tens of GeV, a sufficiently high bulk Lorentz factor at late times can allow a significant flux of gamma rays to escape. Photon-photon absorption due to the EBL can further suppress these gamma rays for GRBs occurring at redshifts beyond the anticipated LIGO/Virgo horizon. An estimate of the rate of detections by CTA of short GRBs associated with GWs gives $\sim 0.03$  yr$^{-1}$~\citep{2014MNRAS.443..738B}; however, considering off-axis events like the recent GRB170817A, this rate should increase~\citep{2017ApJ...848L..13A,2017arXiv171203237L}.

Some long GRBs are associated with the core collapse of massive stars \citep{2006ARA&A..44..507W}. If the collapse is asymmetric enough to produce a detectable GW emission, its signal should precede the burst \citep{2003ApJ...589..861K,2004PhRvD..69d4007V}. {\sl Fermi}-LAT observations showed that very high-energy gamma-ray emission is a relatively common feature in long GRBs (e.g., GRB130427A (\citealt{2014Sci...343...42A}) and GRB080916C (\citealt{2013ApJ...774..76A})). Some very high-energy photons were detected after the prompt emission, such as the 33 GeV photon from GRB090902B, which arrived 82\,s after the trigger time and about 50\,s after the end of the prompt-phase emission (\citealt{2009ApJ...706L.138A}). This burst also exhibited a power-law component at GeV energies,  distinct from the usual Band model emission at MeV energies \citep{1993ApJ...413..281B}, and this component showed significant spectral hardening toward the end of the prompt phase. The theoretical model proposed in \citealt{2014ApJ...788...36B} can explain the emission of GeV photons from long GRBs in connection with their massive progenitors (\citealt{2015ApJ...813...63H}), also predicting TeV emission up to hours after the explosion \citep{2017ApJ...846..152V}.  

The Gamma-ray Burst Monitor (GBM) on the {\sl Fermi} satellite detected a weak transient above 50\,keV, 0.4\,s after the event GW150914, with a false-alarm probability of 2.9$\sigma$ \citep{2016ApJ...826L..6C}. If this transient, lasting 1\,s and with duration and spectrum consistent with a weak short GRB at a large angle to the detector pointing direction, is indeed associated with GW150914 and not a chance coincidence, it is an unexpected electromagnetic emission from a BBH \citep{2016arXiv160207352L}. However, particular environmental conditions for the BBH merger may give sufficient local material to also produce transient high-energy gamma emission  \citep{2013A&A...560A..25J,2016ApJ...819L..21L,2016ApJ...822L...9M,2016ApJ...821L..18P, 2016ApJ...828L...4Z,2016PTEP.2016e1E01Y,2017ApJ...835..165B}. \add{Nonetheless, the GRB origin of Fermi-GBM's detection has been debated \citep{2016arXiv160505447X,2016ApJ...827L..38G}, and} no corresponding signal was found by the \add{SPI instrument on board the} INTEGRAL (\citealt{2016ApJ...820L..36S}) satellite in the same energy region. \add{The AGILE satellite, while it did not cover the GW localization region during the time of GW150914, provided limits on gamma-ray emission in the minutes prior, and following, the prompt event \citep{2016ApJ...825L...4T}.}

Recently, a refined analysis of the data of the MiniCALorimeter (MCAL) on the AGILE satellite, operating in the energy band 0.4-100\,MeV, found a weak event lasting about 32\,ms and occurring 0.46\,s before GW170104, with a \add{post-trial significance of 2.4-2.7$\sigma$} \citep{2017ApJ...847L..20V}, which was also produced by the coalescence of a BBH \citep{2017PhRvL.118v1101A}. The characteristics of this event are similar to those of the weak precursor of short GRB090510, also in its timing, being detected about 0.46\,s before its brightest emission by both AGILE-MCAL and {\sl Fermi}-GBM (\citealt{2010ApJ...708L..84G}, \citealt{2009Natur.462..331A}). If confirmed by different space instruments, this association would prove that a BBH coalescence may be preceded by electromagnetic emission.

\section{Gravitational-Wave Detectors and Observation Strategies}
\label{sec:GW}

\subsection{Gravitational-wave detectors in the CTA era}

In 2018, we expect Advanced LIGO and Virgo to start taking data in their observing run, O3, at almost their design sensitivities, with sensitive ranges to
binary neutron stars between 120-170\,Mpc and 65-85\,Mpc respectively (see Table I in
\citealt{2016LRR....19....1A}). Given the discovery of GW170817, there can potentially be multiple BNS merger detections during this run. By the end of the decade and beyond, advanced detectors will reach their design sensitivities, with ranges to BNS systems of $\sim190\,$Mpc and $\sim125\,$Mpc in Advanced LIGO and Virgo, respectively~\citep{2016LRR....19....1A}. Here, we defined range as the volume and orientation-averaged distance at which the source can be detected. 

Furthermore, there are plans to
implement technology upgrades to Advanced LIGO to further improve
its sensitivity \citep{LIGOPLUS}. This
upgraded Advanced LIGO detector, often referred to as "A+", will likely
come into operation sometime after 2020 and lead to sensitivity corresponding to
a range of 320\,Mpc for BNS signals. 

About 10-20\% of detected BNS events will have sky localization with uncertainties of 20 deg$^2$ or less during O3 when a
three-detector network consisting of the LIGO and Virgo interferometers will
be in operation. With the addition of LIGO India to the global network,
approximately half of all observed BNS events will have sky location
uncertainties of 20 deg$^2$ or less.

The Japanese KAGRA detector is being constructed underground near the Kamioka mines
to reduce seismic noise \citep{PhysRevD.88.043007}. It will have cryogenically cooled test masses
to reduce thermal noise. KAGRA will operate with a simple Michelson
interferometer configuration from 2018 onwards before upgrading to the
full interferometric configuration, with technologies including
Fabry-Perot cavities and signal recycling, in 2019 \citep{2016LRR....19....1A}.

\subsection{Sensitivity to potential gravitational-wave sources}
The detection of GW170817 gives an estimated BNS rate of $1.5^{+3.2}_{-1.2}\times 10^{3}$\, Gpc$^{-3}$yr$^{-1}$ in the local Universe \citep{2017PhRvL.119p1101A}. \add{This rate is remarkably consistent with previous expectations \citep{2010CQGra..27q3001A}. We can use the expected average distance of GW detectors out to which a BNS merger could be detected to calculate the expected detection rate. Assuming (i) an average distance of $120-170$\,Mpc for the O3 observing run and 190\,Mpc at design sensitivity, (ii) 75\% single-detector duty cycle for LIGO and requiring that 2 LIGO detectors are operational for a detection, and (iii) a full year of operation for O3 \citep{2016LRR....19....1A}, we obtain an expected $1-54$ detections for O3, and an expected rate of $5-76$\,yr$^{-1}$ at design sensitivity.}

The lack of NSBH binary detections by Advanced LIGO to date has allowed an upper limit on their rate of 3600 Gpc$^{-3}$yr$^{-1}$~\citep{2016ApJ...832L..21A} to be determined.  

The expected number of NSBH coalescences is $0.01$--$100$ in the O3 observing run and rises in a similar way to the expected BNS coalescences for future observing runs~\citep{2016ApJ...832L..21A}. This range has some sensitivity on the assumptions about the NSBH population, in particular the mass of the black hole and assumptions about the alignment of the black hole spin with the orbital angular momentum. However, going from a conservative population ($5\ M_\odot$ black holes and randomly oriented spins) to an optimistic population ($30\ M_\odot$ black holes with aligned spins) changes the expected rate by only a factor of $3$. For BNS systems we expect spins to be small and the two components to have masses close to $1.4\ M_\odot$ so there is little dependence on the system parameters. 

Advanced LIGO observations of GW150914, LVT151012 and GW151226 have constrained the rate of binary black hole coalescences to be in the range $12$--$213$ Gpc$^{-3}$yr$^{-1}$~\citep{2017PhRvL.118v1101A}. This range includes both statistical uncertainties and uncertainties arising from the astrophysical population of BBH systems. The higher end of the rate range comes from assuming that the mass distribution in BBH binaries follows a power-law with slope $\alpha = -2.35$, and the lower end comes from assuming the mass distribution is flat in logarithmic scale. The power-law distribution intrinsically predicts fewer heavier black hole systems, to which Advanced LIGO is more sensitive, and hence the allowed rate of mergers is higher since it is dominated by systems that Advanced LIGO can only detect at moderate distances~\citep{2016ApJ...833L...1A}. During the O3 science run, Advanced LIGO should be able to detect equal-mass, non-spinning BBH systems with total mass of $20M_\odot$ out to distances of $\sim 1.5$\,Gpc. The range increases as the total mass of the system increases, to $\sim 3$\,Gpc for systems of total mass $40\,$M$_\odot$, to $\sim 4.5$\,Gpc for systems with total mass of $60$\,M$_\odot$ and then $\sim 8$\,Gpc for systems with total mass of $100$\,M$_\odot$~\citep{2016ApJ...818L..22A}. All of these ranges approximately double by the time Advanced LIGO achieves its design sensitivity. The range is increased if the components in the BBH  have significant spins, and reduced (at fixed total mass) if the two components have unequal masses. The events observed by Advanced LIGO/Virgo to date are all nearly equal mass (mass ratios between $0.5$ and $1$) and only GW151226 shows evidence for spin, and that spin is moderate (effective spin $\sim 0.2$)~\citep{2016PhRvX...6d1015A}. If these events are representative of the true astrophysical population, the non-spinning equal-mass ranges provide an accurate indication of LIGO's likely sensitivity to BBH mergers.

Based on current rate estimates, there is between a $90\%$ and $99\%$ probability that Advanced LIGO and Virgo will observe more than $10$ BBH mergers during the O3 science run in 2018--2019. The probability range arises from uncertainties in the sensitivity that will be achieved during that science run. There is also a probability of between $20\%$ and $80\%$ that Advanced LIGO and Virgo will observe more than $40$ BBH events during O3~\citep{2016PhRvX...6d1015A}. At design sensitivity, Advanced LIGO's range will be double that during O3 and so it will be probing a comoving volume between $3$ and $5$ times larger (depending on the mass of the BBH system). We would therefore expect Advanced LIGO to be observing one hundred events per year at design sensitivity with high probability, with a plausible range for the expected number of detections per year of $\sim 50$--$1000$~\citep{2016ApJ...833L...1A,2016ApJS..227...14A,2016PhRvX...6d1015A}.

Numerical CCSN simulations suggest that Advanced LIGO/Virgo may be able to detect GWs from a CCSN event if it will occur within the Milky Way, but probably not at greater distances. However, if GW emission will be generated at much higher amplitudes than numerical simulations suggest, as predicted by some extreme analytic models of exotic emission scenarios, then Advanced LIGO could have a distance reach of as much as $\sim10$\,Mpc~\citep{2016PhRvD..94j2001A}. The Galactic supernova rate is $0.02-0.03/$\,yr, which makes a detection during Advanced LIGO/Virgo operations plausible but improbable, although a Milky Way supernova is overdue. The CCSN rate within $15$\,Mpc is $\sim1/$\,yr, so the prospects for Advanced LIGO detection are significantly improved if CCSNe do generate much more GW emission than current numerical models suggest~\citep{2016PhRvD..94j2001A}.

\subsection{Low-latency electromagnetic follow-up observations}

The utility of GWs in studying astrophysical processes is greatly increased by the simultaneous observation of electromagnetic and/or neutrino emission from the same sources. Consequently, there is a significant effort to enable Earth-based GW detectors to rapidly identify and localize GW source candidates, and share this information with partner observatories \citep{2007AAS...211.9903P,2008CQGra..25r4034K,2012A&A...539A.124L,2013APh....45...56S,2016ApJ...826L..13A,2016PhRvD..93l2010A}. GW candidates were rapidly shared with a large number of partner observatories already during the operation of Initial LIGO-Virgo \citep{2012A&A...541A.155A}, which was further expanded during LIGO's first observing run (O1; \citealt{2016ApJ...826L..13A}). \add{Additional improvements and an increased quality of communication were implemented for the second observing run (O2; \citealt{2017ApJ...848L..12A}).}

Going forward, GW candidates will be shared at increasing rates and decreasing latency. The false alarm rate of shared triggers was set at 1/month during the O2 observing run both for compact binary merger candidates and separately for GW transient candidates identified with minimal assumptions on the source \citep{2016PhRvD..93l2004A}. The trigger rate will likely be higher than this, given the expected large and growing rate of GW detections \citep{2016arXiv160604856T}.

Follow-up observatories will be able to selectively investigate GW triggers. For example, they can constrain follow-up searches based on the reconstructed progenitor type, source distance, direction, significance and other parameters. This sub-selection can optimize the use of telescope time and the follow-up strategy.

A significant amount of information will be rapidly available from GW reconstruction algorithms that can facilitate optimizing follow-up strategies. Information on GW candidates include their time of arrival, localization (skymap), the type of GW search pipeline that detected the candidate (template-based compact binary search, generic transient search with minimal assumptions, or both), and false alarm rate. Additionally, for compact binary mergers, a source distance estimate will be available, along with whether at least one of the objects in the binary is a neutron star that could be disrupted. This latter case differentiates between likely electromagnetically bright sources, such as (i) binary neutron stars or (ii) black hole--neutron star pairs with relatively small black hole mass such that the neutron star may be disrupted, from (i) binary black holes or (ii) black hole--neutron star pairs with larger black hole mass such that the neutron star plunges into the black hole without disruption. For GW transient candidates other than binary mergers, the observed duration, characteristic frequency, and total GW fluence will be available and can help constrain the possible source types and distances.

GW candidates are currently identified with a latency of $\sim1$\,min. Following this initial detection, candidates undergo manual checks (by humans) to ensure data quality, which introduces an additional $\sim 30$\,min delay. For some follow-up observatories, such as CTA, humans in the loop introduce too long a delay given the expected short duration of high-energy emission, and these observatories will therefore need to rely on the earliest available reconstruction.

\section{The Cherenkov Telescope Array and Observation Strategies}
\label{sec:CTA}

In terms of raw sensitivity, CTA can detect GRBs to very high redshifts as long as the GRB spectrum continues out to very high energies when extrapolating from the observed {\sl Fermi}-LAT spectrum (e.g., \citealt{2013APh....43..252I}). However, the observed spectra at very high energies will be attenuated by the interaction of source photons with the EBL \citep{2001ARA&A..39..249H,2011MNRAS.410.2556D}. In the redshift range probed by Advanced LIGO and Virgo the attenuation will be small below a few hundred GeV (e.g., \citealt{2013ApJ...768..197I}), which is the most likely energy range for CTA to detect a burst.

\subsection{CTA telescopes}

The CTA observatory is being designed by an international consortium, which is currently building prototypes and characterizing them\footnote{\url{https://www.cta-observatory.org}}. To provide all-sky access, CTA will comprise two arrays, with one deployed in the Northern hemisphere, on La Palma (Spain), while Paranal (Chile) is the site in the Southern hemisphere. Meeting the ambitious CTA design goals, including an overall increase in sensitivity of about an order of magnitude compared with the current generation of IACTs (\citealt{2017arXiv170907997C}), requires a large number of telescopes of different sizes in order to cover the energy range from   20\,GeV up to above 100\,TeV. The telescopes are grouped in three sizes; the large-sized (23\,m diameter\ \textbf{LSTs}), medium-sized (12\,m \textbf{MSTs}) and small-sized (4\,m \textbf{SSTs}). A prototype of a dual-mirror version of the MST (Schwarzschild-Couder Telescope (\textbf{SCT}) with a 9.7-m primary mirror) is also being built. The LSTs provide access to the low-energy range ($\leq0.1$\,TeV), the SSTs to the high-energy range ($>10$\,TeV) while the MSTs ensure enhanced sensitivity in the core energy range of CTA ($0.1 - 10$\,TeV). The telescopes will be arranged on the ground such that the LSTs are grouped together, aiming to sample a substantial fraction ($\approx 10\%$) of the Cherenkov light pool, surrounded by an array of MSTs, that will ensure an excellent shower reconstruction due to a large stereoscopic multiplicity, and a more numerous collection of SSTs that will extend the array footprint, thus increasing the effective area of the instrument in a domain where event statistics are the main limiting factor. Going from the lowest to the highest energies, CTA will provide an angular resolution from $\sim0.25$\textdegree\ to $\sim0.03$\textdegree\ and a field of view from $\sim5$\textdegree\ to $\sim10$\textdegree \citep{2017arXiv170907997C}. However, the sensitivity across the field of view is not uniform, particularly below 100 GeV, where the sensitivity is dominated by the LSTs and is $\sim50$\% lower at 1.5\textdegree\ off-axis than it is on axis \citep{2017arXiv170901381M}. At the lowest energies, \add{where there is some overlap of CTA and {\sl Fermi}-LAT, the former gains the most in sensitivity on short timescales compared to the latter \citep{2013APh....43..348F}, which on the other hand has a greater sky coverage. Thus, slew speed is important, and} the LSTs are designed to slew to any point in the sky within 20\,s, while the MSTs can slew to any point in 90\,s.

\subsection{Array Deployment}

The array baseline design proposed to provide the required sensitivity and energy range is for 4 LSTs and 15 MSTs in the Northern hemisphere and 4 LSTs, 25 MSTs and 70 SSTs in the Southern hemisphere. The SSTs will be deployed only in the South to enhance observations of Galactic plane sources. Concerning the deployment schedule, the pre-production phase calls for the first CTA telescopes to be installed in 2019, which will include a prototype LST on La Palma, at the site where the two Major Atmospheric Gamma-ray Imaging Cherenkov (MAGIC) telescopes are currently hosted \citep{2016APh....43...76A}. The production phase follows until full commissioning and science verification is completed. The field of view of a given size telescope does not vary significantly, but sensitivity and angular resolution do improve as more telescopes are added into the array. Transient observations will start early on during construction (in principle with only a single telescope in operation), but initially, due to the current LST deployment schedule, the lowest energy range of CTA will be available only from the North.

\subsection{Transient Follow-up}

GW transients are proposed as the highest priority target to be investigated by CTA with a rapid and coordinated response within CTA's Key Science Project on transients. During its early phase, before array completion, a maximum of 20\,h\,yr$^{-1}$ of observation time is foreseen for each CTA site to follow up GW transients, reduced to 5-10\,h\,yr$^{-1}$ in the subsequent years with the full array \citep{2017arXiv170907997C}. CTA has some unique capabilities for GW follow-up, including a large instantaneous field of view and access to both hemispheres \citep{2014MNRAS.443..738B}. As GW alerts can have large associated localization regions, both the northern and southern CTA arrays may have to be triggered to ensure coverage. The field of view of the CTA telescopes is sufficiently large that some galaxies within the  Advanced LIGO/VIRGO horizon should always be within view and hence targeting could be based only on the localization probability map via a set of pointing directions (tiling).

In \citealt{2014MNRAS.443..738B} it was demonstrated that CTA will be capable of following up GW transients associated to short GRBs inside the horizon distance of Advanced LIGO/Virgo over a sky area as large as $\sim$1000\,deg$^2$. Considering here the detectability of a transient gamma-ray source with unknown direction (survey mode) over an area of $\sim$100 deg$^2$, we have to take into account the scaling factor $f_\Omega \simeq [(\theta_{CTA}/2)^2 \pi/\Omega_{GW}]^{1/2}$, where $\theta_{CTA}$ is the diameter of the field of view of CTA and $\Omega_{GW}$ is the area of the error region of the GW transient. Thus, the fluence threshold is reduced by a factor $\approx 3$ compared to \cite{2014MNRAS.443..738B}, increasing the detectability of intrinsically weaker sources, as well as sources with a cutoff at lower energies and/or with a larger delay in the start of observations after the trigger (see Table 1 in \citealt{2014MNRAS.443..738B}) and/or larger zenith angles. 

For a GeV gamma-ray source with a temporal decay $\propto t^{-1.38}$, like that observed in the short GRB090510 \citep{2010ApJ...709L.146D}, an emission lasting 100\,s in an observation starting 100\,s after the trigger gives a fluence lower by a factor $\approx 3$ with respect to a duration of 1000\,s, as considered in \citealt{2014MNRAS.443..738B}. Therefore, the improvement in the sky localization of the GW transients from $\sim$1000 deg$^2$ to $\sim$100 deg$^2$ increases also the detectability of sources with shorter duration, of the order of the observations given in \citealt{2010ApJ...709L.146D}. 

Alternative observation modes with CTA, like the so-called "divergent" pointing mode, are under study~\citep{2015APh....67...33S}. Exploratory simulations of single-dish MST arrays indicate that effective fields of view of $20$\textdegree\ can be achieved from divergent pointing, offering a better sensitivity than the conventional parallel pointing when scanning large portions of the sky (at the expense of energy and angular resolutions). If proved efficient, this observation mode could be employed for the follow-up of GW transients, reducing the time required to tile a large localization region at a given sensitivity. If useful for surveys, this operational mode would also increase the probability of a joint prompt detection for short-duration transients. For a given target sensitivity, a further reduction to the time required to survey the localization region may come from the better off-axis performance provided by the SCTs as compared to single-dished MSTs \citep{Hassan:2015yva}, at the price of a slightly increased energy threshold. 

The Real-Time Analysis (RTA) pipeline will automatically determine if a new source has been detected and issue an alert within 30 seconds from the triggering event collection, ensuring fast communication with the astrophysics community (e.g., with Virtual Observatory Events) (\citealt{2015arXiv150901943F}).
On short timescales (hundreds of seconds), CTA is unlikely to detect a new steady source, but in any case CTA follow-up observations will occur for any new source detected. The CTA design requires that the sensitivity of the RTA search for transients (on multiple time scales) should be not worse than three times the nominal CTA performance. Assuming a factor three, the RTA with the southern full array will achieve in a few hours of observations and within 30 seconds of reconstruction and processing the same sensitivity of 50 hours of integration of the current Cherenkov telescopes. As a reference value, the Southern CTA array will be able to detect 10\% of the Crab Nebula integral flux with 1000 seconds of pointed observation, for an energy threshold less than 10\,TeV \citep{2016SPIE.9906E.3OF}.
However, there is still space for improvements in the algorithms and hardware, and thus in the decrease of the minimum detectable flux with the RTA. 

The CTA Consortium will receive GW alerts from the Advanced LIGO/Virgo interferometers as stipulated in a memorandum of understanding already signed, and will follow-up those during dark time with zenith angles less than 70\textdegree\ for 2 hours each, adding exposure time in case of positive detections \citep{2017arXiv170907997C}. 

The duty cycle of current-generation IACTs is affected, among other factors, by the lunar phase, which prevents observations during full Moon due to the elevated brightness of the sky, thus potentially reducing the overlap between \add{Advanced} LIGO/Virgo and present IACT uptimes. However, observations are routinely performed under moderate moonlight, representing a $\sim30\%$ increase over an average, dark-sky only, observing year (e.g., \citealt{Archambault201734}). SiPM-based IACT cameras have proven to be effective in the detection of cosmic showers under bright moonlight conditions without risking the photodetectors' integrity or accelerating their aging \citep{2014JInst...9P0012B}, allowing for an increased duty cycle, although with reduced sensitivity and larger energy threshold. This technological advancement will be utilized in the SCT camera \citep{Otte:2015mya}, as well as in the SST cameras \citep{Montaruli:2015xya}, therefore opening the possibility of following up GW alerts even during bright moonlight conditions.  

The closeness of the MAGIC telescopes and the prototype LST may give the opportunity, if they will be operated in coincidence, to start carrying out a follow-up of GW transients at the CTA Northern site with a system of three large and fast slewing Cherenkov telescopes in stereoscopic mode. 

\section{Joint Search Methodology}
\label{ref:jointsearch}

\subsection{Previous Search Strategies With Cherenkov Telescopes}

The current-generation IACTs -- H.E.S.S., MAGIC, and VERITAS -- have been used to perform searches of very high-energy gamma-ray emission associated with GW triggers. The cameras of these IACTs cover, with radially-dependent sensitivity, an area in the sky with a size between $\sim8$ and $\sim20$ square degrees, making them suitable to survey a fraction of the localization uncertainty regions of LIGO and LIGO/Virgo events. We briefly discuss results from these searches as they provide a learning experience for future follow-up observations using CTA.

During the O1 run, MAGIC performed follow-up observations~\citep{Carosi:2017tbk} of the event GW151226, later identified as due to a BBH merger~\citep{PhysRevLett.116.241103}. The event was detected on 2015 December 26 UT and a localization map was circulated by LIGO the following day, based on which four MAGIC pointings were manually selected maximizing the probability coverage and taking into account visibility, overlap with existing catalogs, and observations of other telescopes. The four positions were observed starting on December 28 UT with an average exposure of 42 min per pointing\footnote{https://gcn.gsfc.nasa.gov/gcn3/18776.gcn3}. No source was detected during these observations. 

The first VHE follow-up  during the O2 run was performed by VERITAS~\citep{Santander:2016chw} for the event GW170104~\citep{2017PhRvL.118v1101A} detected on 2017 January 4 UT which was due to the coalescence of a 50-Solar-mass BBH system at a redshift of 0.2. VERITAS opted for tiling the Northern fraction of the localization map above a 50\textdegree\ elevation using 39 consecutive pointings each observed for approximately five minutes. The survey started on January 5 UT and covered 27 \% of the event containment probability. Although the presence of clouds affected observations, VERITAS reports that these observations were sensitive to sources with a flux greater than 50 \% of the Crab nebula above 100 GeV\footnote{https://gcn.gsfc.nasa.gov/gcn3/21153.gcn3}.

The first detection of gravitational waves from binary neutron stars (GW170817) took place during O2. IACT observations were started by H.E.S.S. 5.3 h after the detection of the event using an observational strategy that identified regions of high probability to find a GW counterpart. The first of these observed regions included the location of SSS17a, the EM counterpart for GW170817 identified later in the optical range. Two algorithms developed by H.E.S.S. optimized for real-time GW follow up and offline scheduling are detailed in~\cite{2017arXiv170510138S} which include folding the localization maps for the GW events with a galaxy catalog, and the prioritization of different targets according to their distribution in the sky and observational constraints. For this follow-up, observations were started on Aug 17-18 UT and continued over several days, setting upper limits in the energy band between 0.28-8.55 TeV as no gamma-ray excess was identified in the observed region~\citep{2017arXiv171005862H}. 

\subsection{LIGO/Virgo Alerts and Follow-up}

During the O2 observing period, LIGO and Virgo sent out alerts typically within about 30 minutes of the identification of an interesting event; most of the time taken to generate alerts was due to human vetting of identified events; as a case study of the alert process, we briefly summarize the steps leading up to the publication of the GCN alert that notified other observatories of the detection and likely source direction. We use the binary neutron star event GW170817 as our case.

LIGO recorded the GW event in data from its Hanford detector with a low-latency search 6 minutes after the merger \citep{2017PhRvL.119p1101A,2017ApJ...848L..12A}. LIGO-Livingston data was initially not used due to a noise artifact overlapping with the signal, which was later removed. Virgo also recorded the signal, but its low-latency data transfer was delayed. LIGO and Virgo sent a GCN alert about 35 minutes after the BNS event was registered, corresponding to about 40 minutes after the BNS merger. A GW skymap using information from both LIGO detectors and Virgo was distributed five hours after the merger. Compared to the {\sl Fermi}-GBM skymap for GRB170817A, this skymap identified a smaller region of the sky and led to the discovery of the optical transient and host galaxy identification. For comparison, the {\sl Fermi}-GBM detection and initial skymap were distributed 14 seconds after detection \citep{2017ApJ...848L..14G}.

We see that, while the analysis and the dissemination of information from GW170817 was too long for practical use by high-energy observatories like CTA, most delay occurred due to human involvement and technical difficulties that can be overcome for future detections.

Looking at the end result, the reconstructed $90\%$ CL skymap of GW170817 was only $\sim30$ deg$^2$. This localization and the source distance are much more favorable than the more conservative case discussed by \cite{2014MNRAS.443..738B}, making the detection prospects of CTA promising, given that the delay due to GW data analysis is comparable to the 6 minutes achieved for GW170817. Note that, for the particular case of GW170817 in which detection was established early but the GW skymap was significantly delayed, it can also be feasible to take the fact of GW observation and the localization from the corresponding GRB, and scan the corresponding sky area \add{(in fact this is what happened with GW170817--observers scanned the Fermi localization region until an improved GW localization area became available)}.

More generally, GW detector networks can typically give larger sky localization regions (see \citealt{2014ApJ...795..105S} for details). In these cases, CTA will need to cover as much of the GW sky localization region as possible. There have been multiple studies discussing optimal strategies for covering the broad GW sky localization regions and maximizing the probability of detecting transient counterparts (e.g., \citealt{2017ApJ...834...84C,2016ExA....42..165C,2016A&A...592A..82G,2017ApJ...846...62S}). All of these studies highlight that the probability of detecting a counterpart transient can be boosted by factors of a few if the sky location is tiled efficiently and the time allocation per sky location is optimized for each telescope's sensitivity and the skymap probability. 

Additionally, the probability of counterpart detection can be boosted if observing strategies target galaxies within the GW sky localization region \citep{2017ApJ...834...84C,2016ApJ...820..136G,2016ApJ...829L..15S}. A good example for this strategy is that followed by the Swope Telescope which was the first to discover the optical counterpart of GW170817 despite its small telescope size \citep{2017arXiv171005452C}.

It is relevant to note here that rapid skymaps generated for interesting GW events do not incorporate calibration uncertainties in GW detectors. Such calibrations are generated only after several days. We also note that the probability assigned to different regions in GW skymaps can change as the skymap is refined (see, e.g., \citealt{2016PhRvL.116f1102A}). 

An interesting future possibility is the first detection of gravitational waves from a phenomenon other than compact binary mergers. For such a case, due to limited signal models, there could be additional uncertainties associated with the skymap. At the same time, such sources are expected to be detectable only at smaller distances than compact binary sources, making their potential observable very high-energy emission brighter and more detectable. Despite the large size of GW sky localization regions, CTA will typically have the capacity to cover the GW sky region even for these cases.

Since for most foreseeable cases, CTA will be able to cover the reconstructed GW skymap, and since these skymaps are typically generated at 90\% confidence level, upon non detection it will be beneficial to allocate remaining observation resources to observe regions around the estimated GW sky localization, as contingency against variations in the GW sky localization due to factors laid out above.

\section{Conclusion and Outlook}
\label{sec:conclusion}

We reviewed the current status of GW-multimessenger observations and that of CTA, in order to examine the possible strategies of searching for very high-energy emission from GW transients with CTA. Our main goals were to present a summary to the GW and CTA communities of progress on the other side, as well as to identify what needs to be done before joint observations commence. This work was a continuation of \cite{2014MNRAS.443..738B}, where we explored the utility of CTA for GW follow-up observations, and found that CTA is well-suited to work with even large localization uncertainties, often expected for GW signals.

Based on the present status of observatories and multimessenger observations, we consider the following directions to be important to maximize the GW-follow-up potential of CTA:
\begin{enumerate}
\item {\bf Low-latency alerts:} With the rapid fading of high-energy emission in the aftermath of a binary merger, it is critical that GW detections are shared with low latency with partner observatories. This requires full automation on the GW side.  For CTA, a higher false alarm rate is tolerable as the follow-up of GW triggers only requires $O(1000\mbox{s})$ of observation time. Additionally, automation in the reception and response to the alert by CTA, and in the execution of the follow-up observation, is also necessary to facilitate an optimal outcome.
\item {\bf Start with a single CTA telescope:} If gamma-ray emission from short GRBs extends to E $>\,$20 GeV, CTA may be able to detect such emission even with a single LST on $\sim100$\,Mpc distance scales relevant for GW observations. The commissioning of an automated alert system and the corresponding CTA follow-up observation execution should begin as soon as a single CTA telescope is deployed. 
\item {\bf No need for galaxy catalogs:} While galaxy catalogs played an important role in the follow-up of GW170817, their utility for CTA follow-ups will be limited. Since CTA will have a large multi-deg$^2$ field of view, albeit with some sensitivity degradation off axis, a non-uniform galaxy distribution will rarely impact the prioritization of pointing directions. Additionally, the very high-energy sky has few transients, therefore galaxy catalogs are not needed to reduce the false-alarm rate. 
\item {\bf Most GW candidates can be followed up:} With CTA's dedicated GW-follow-up observing time of $\sim10$\,hr\,yr$^{-1}$, considering an observation time of $1000$\,s per event \citep{2014MNRAS.443..738B}, we expect that CTA will be able to follow-up all GW candidates other than binary black hole mergers that fall within the region of the sky accessible to CTA, given the expected rate of non-BBH GW candidates of one per month, and a one-per-month false alarm rate. This means that no prioritization is needed based on the properties of the non-BBH GW candidates, allowing the potential for the discovery of unusual sources. For BBH mergers, the detection rate could be several hundred per year once LIGO/Virgo reach their design sensitivity \citep{2016LRR....19....1A,2016PhRvX...6d1015A}. Such a high rate will be unfeasible to comprehensively follow up and some prioritization will be needed.
\item {\bf Deeper observation of promising events:} Some of the available observing time should be utilized for a deeper observation of a promising GW event, preferably extending CTA's original observation, in so far as this is technically viable, rather than observing the region of interest in successive nights. Such an extension can be motivated by insight in the nature of the event. For instance, such an event can be a binary neutron star merger whose reconstructed parameters indicate that it is nearby and its orbital axis is roughly pointing towards Earth, or even with an observed GRB counterpart. For such an event, if an initial scan with CTA finds no very high-energy emission, it is worth integrating for longer in order to enable probing the high-energy cutoff of gamma-ray emission from the event. A complementary motivation for a deeper observation is the prompt finding of a hint of signal in CTA's RTA.
\item {\bf Multi-messenger follow-up:} Cosmic messengers that are promptly emitted from GW sources and that are monitored by ``all-sky" detectors can be rapidly available along with GWs. Such messengers include gamma-rays and high-energy neutrinos. It will be useful to plan CTA observations such that the low-latency detection of a GRB counterpart, or high-energy neutrinos, from a GW source can be incorporated in the follow-up. In particular, these other messengers can significantly improve the localization of the source. For instance, high-energy neutrino track events can be reconstructed to sub-degree precision \citep{2017ApJ...850L..35A}.
\item {\bf Multi-messenger alert:} Once CTA identifies very high-energy emission from a GW source, its precise direction reconstruction ($\lesssim 0.1$\textdegree; \citealt{2013APh....43..171B, 2017arXiv170907997C}) can be used to point other follow-up observatories in the right source direction. It is important that such identification is communicated to partner observatories as soon as possible. For example, X-ray emission may rapidly fade similarly to very high-energy emission, and with the narrow fields of view of the most sensitive current instruments it will be beneficial to learn the true source direction quickly.
\end{enumerate}

\section*{Acknowledgments}

The authors thank Giulia Stratta for useful discussions. This paper has gone through internal review and has been approved for publication by the LIGO Scientific Collaboration, the Virgo Collaboration and the CTA Consortium. IB, SM, ZM and TBH are thankful for the generous support of Columbia University in the City of New York. IB, ZM and SM are thankful for the generous support of the National Science Foundation under cooperative agreement PHY-1708028. IB is thankful for the generous support of the University of Florida. TBH is thankful for the generous support of the National Science Foundation under cooperative agreement PHY-1352567. DN is thankful for the generous support of the Spanish Ministry of Economy, Industry and Competitiveness, and the European Regional Development Fund through the grant FPA2015-73913-JIN. TDG is thankful for the kind hospitality of Columbia University in the City of New York.


\bibliographystyle{mnras}

\begin{thebibliography}{}
\makeatletter
\relax
\def\mn@urlcharsother{\let\do\@makeother \do\$\do\&\do\#\do\^\do\_\do\%\do\~}
\def\mn@doi{\begingroup\mn@urlcharsother \@ifnextchar [ {\mn@doi@}
  {\mn@doi@[]}}
\def\mn@doi@[#1]#2{\def\@tempa{#1}\ifx\@tempa\@empty \href
  {http://dx.doi.org/#2} {doi:#2}\else \href {http://dx.doi.org/#2} {#1}\fi
  \endgroup}
\def\mn@eprint#1#2{\mn@eprint@#1:#2::\@nil}
\def\mn@eprint@arXiv#1{\href {http://arxiv.org/abs/#1} {{\tt arXiv:#1}}}
\def\mn@eprint@dblp#1{\href {http://dblp.uni-trier.de/rec/bibtex/#1.xml}
  {dblp:#1}}
\def\mn@eprint@#1:#2:#3:#4\@nil{\def\@tempa {#1}\def\@tempb {#2}\def\@tempc
  {#3}\ifx \@tempc \@empty \let \@tempc \@tempb \let \@tempb \@tempa \fi \ifx
  \@tempb \@empty \def\@tempb {arXiv}\fi \@ifundefined
  {mn@eprint@\@tempb}{\@tempb:\@tempc}{\expandafter \expandafter \csname
  mn@eprint@\@tempb\endcsname \expandafter{\@tempc}}}

\bibitem[\protect\citeauthoryear{{Aasi} et~al.,}{{Aasi}
  et~al.}{2014}]{2014PhRvD..89j2006A}
{Aasi} J.,  et~al., 2014, \mn@doi [\prd] {10.1103/PhysRevD.89.102006}, \href
  {http://adsabs.harvard.edu/abs/2014PhRvD..89j2006A} {89, 102006}

\bibitem[\protect\citeauthoryear{{Abadie} et~al.,}{{Abadie}
  et~al.}{2010}]{2010CQGra..27q3001A}
{Abadie} J.,  et~al., 2010, \mn@doi [Classical and Quantum Gravity]
  {10.1088/0264-9381/27/17/173001}, \href
  {http://adsabs.harvard.edu/abs/2010CQGra..27q3001A} {27, 173001}

\bibitem[\protect\citeauthoryear{{Abadie} et~al.,}{{Abadie}
  et~al.}{2012}]{2012A&A...541A.155A}
{Abadie} J.,  et~al., 2012, \mn@doi [\aap] {10.1051/0004-6361/201218860}, \href
  {http://adsabs.harvard.edu/abs/2012A%26A...541A.155A} {541, A155}

\bibitem[\protect\citeauthoryear{{Abbott} et~al.,}{{Abbott}
  et~al.}{2016a}]{2016PhRvX...6d1015A}
{Abbott} B.~P.,  et~al., 2016a, \mn@doi [Physical Review X]
  {10.1103/PhysRevX.6.041015}, \href
  {http://adsabs.harvard.edu/abs/2016PhRvX...6d1015A} {6, 041015}

\bibitem[\protect\citeauthoryear{{Abbott} et~al.,}{{Abbott}
  et~al.}{2016b}]{2016LRR....19....1A}
{Abbott} B.~P.,  et~al., 2016b, \mn@doi [Living Reviews in Relativity]
  {10.1007/lrr-2016-1}, \href
  {http://adsabs.harvard.edu/abs/2016LRR....19....1A} {19}

\bibitem[\protect\citeauthoryear{{Abbott} et~al.,}{{Abbott}
  et~al.}{2016c}]{2016PhRvD..93l2004A}
{Abbott} B.~P.,  et~al., 2016c, \mn@doi [\prd] {10.1103/PhysRevD.93.122004},
  \href {http://adsabs.harvard.edu/abs/2016PhRvD..93l2004A} {93, 122004}

\bibitem[\protect\citeauthoryear{{Abbott} et~al.,}{{Abbott}
  et~al.}{2016d}]{2016PhRvD..94j2001A}
{Abbott} B.~P.,  et~al., 2016d, \mn@doi [\prd] {10.1103/PhysRevD.94.102001},
  \href {http://adsabs.harvard.edu/abs/2016PhRvD..94j2001A} {94, 102001}

\bibitem[\protect\citeauthoryear{{Abbott} et~al.,}{{Abbott}
  et~al.}{2016e}]{2016PhRvL.116f1102A}
{Abbott} B.~P.,  et~al., 2016e, \mn@doi [Physical Review Letters]
  {10.1103/PhysRevLett.116.061102}, \href
  {http://adsabs.harvard.edu/abs/2016PhRvL.116f1102A} {116, 061102}

\bibitem[\protect\citeauthoryear{Abbott et~al.}{Abbott
  et~al.}{2016f}]{PhysRevLett.116.241103}
Abbott B.~P.,  et~al., 2016f, \mn@doi [Physical Review Letters]
  {10.1103/PhysRevLett.116.241103}, \href
  {http://adsabs.harvard.edu/abs/2016PhRvL.116x1103A} {116, 241103}

\bibitem[\protect\citeauthoryear{{Abbott} et~al.,}{{Abbott}
  et~al.}{2016g}]{2016ApJS..227...14A}
{Abbott} B.~P.,  et~al., 2016g, \mn@doi [\apjs] {10.3847/0067-0049/227/2/14},
  \href {http://adsabs.harvard.edu/abs/2016ApJS..227...14A} {227, 14}

\bibitem[\protect\citeauthoryear{{Abbott} et~al.,}{{Abbott}
  et~al.}{2016h}]{2016ApJ...818L..22A}
{Abbott} B.~P.,  et~al., 2016h, \mn@doi [\apjl] {10.3847/2041-8205/818/2/L22},
  \href {http://adsabs.harvard.edu/abs/2016ApJ...818L..22A} {818, L22}

\bibitem[\protect\citeauthoryear{{Abbott} et~al.,}{{Abbott}
  et~al.}{2016i}]{2016ApJ...826L..13A}
{Abbott} B.~P.,  et~al., 2016i, \mn@doi [\apjl] {10.3847/2041-8205/826/1/L13},
  \href {http://adsabs.harvard.edu/abs/2016ApJ...826L..13A} {826, L13}

\bibitem[\protect\citeauthoryear{{Abbott} et~al.,}{{Abbott}
  et~al.}{2016j}]{2016ApJ...832L..21A}
{Abbott} B.~P.,  et~al., 2016j, \mn@doi [\apjl] {10.3847/2041-8205/832/2/L21},
  \href {http://adsabs.harvard.edu/abs/2016ApJ...832L..21A} {832, L21}

\bibitem[\protect\citeauthoryear{{Abbott} et~al.,}{{Abbott}
  et~al.}{2016k}]{2016ApJ...833L...1A}
{Abbott} B.~P.,  et~al., 2016k, \mn@doi [\apjl] {10.3847/2041-8205/833/1/L1},
  \href {http://adsabs.harvard.edu/abs/2016ApJ...833L...1A} {833, L1}

\bibitem[\protect\citeauthoryear{{Abbott} et~al.,}{{Abbott}
  et~al.}{2017a}]{2017PhRvL.118v1101A}
{Abbott} B.~P.,  et~al., 2017a, \mn@doi [Physical Review Letters]
  {10.1103/PhysRevLett.118.221101}, \href
  {http://adsabs.harvard.edu/abs/2017PhRvL.118v1101A} {118, 221101}

\bibitem[\protect\citeauthoryear{{Abbott} et~al.,}{{Abbott}
  et~al.}{2017b}]{2017PhRvL.119p1101A}
{Abbott} B.~P.,  et~al., 2017b, \mn@doi [Physical Review Letters]
  {10.1103/PhysRevLett.119.161101}, \href
  {http://adsabs.harvard.edu/abs/2017PhRvL.119p1101A} {119, 161101}

\bibitem[\protect\citeauthoryear{{Abbott} et~al.,}{{Abbott}
  et~al.}{2017c}]{2017ApJ...848L..12A}
{Abbott} B.~P.,  et~al., 2017c, \mn@doi [\apjl] {10.3847/2041-8213/aa91c9},
  \href {http://adsabs.harvard.edu/abs/2017ApJ...848L..12A} {848, L12}

\bibitem[\protect\citeauthoryear{{Abbott} et~al.,}{{Abbott}
  et~al.}{2017d}]{2017ApJ...848L..13A}
{Abbott} B.~P.,  et~al., 2017d, \mn@doi [\apjl] {10.3847/2041-8213/aa920c},
  \href {http://adsabs.harvard.edu/abs/2017ApJ...848L..13A} {848, L13}

\bibitem[\protect\citeauthoryear{{Abdalla} et~al.,}{{Abdalla}
  et~al.}{2017}]{2017arXiv171005862H}
{Abdalla} H.,  et~al., 2017, preprint, \href
  {http://adsabs.harvard.edu/abs/2017arXiv171005862H} {} (\mn@eprint {arXiv}
  {1710.05862})

\bibitem[\protect\citeauthoryear{{Abdo} et~al.,}{{Abdo}
  et~al.}{2009a}]{2009Natur.462..331A}
{Abdo} A.~A.,  et~al., 2009a, \mn@doi [\nat] {10.1038/nature08574}, \href
  {http://adsabs.harvard.edu/abs/2009Natur.462..331A} {462, 331}

\bibitem[\protect\citeauthoryear{{Abdo} et~al.,}{{Abdo}
  et~al.}{2009b}]{2009ApJ...706L.138A}
{Abdo} A.~A.,  et~al., 2009b, \mn@doi [\apjl] {10.1088/0004-637X/706/1/L138},
  \href {http://adsabs.harvard.edu/abs/2009ApJ...706L.138A} {706, L138}

\bibitem[\protect\citeauthoryear{{Acciari} et~al.,}{{Acciari}
  et~al.}{2011}]{2011ApJ...743...62A}
{Acciari} V.~A.,  et~al., 2011, \mn@doi [\apj] {10.1088/0004-637X/743/1/62},
  \href {http://adsabs.harvard.edu/abs/2011ApJ...743...62A} {743, 62}

\bibitem[\protect\citeauthoryear{{Acernese} et~al.,}{{Acernese}
  et~al.}{2015}]{2015CQGra..32b4001A}
{Acernese} F.,  et~al., 2015, \mn@doi [Classical and Quantum Gravity]
  {10.1088/0264-9381/32/2/024001}, \href
  {http://adsabs.harvard.edu/abs/2015CQGra..32b4001A} {32, 024001}

\bibitem[\protect\citeauthoryear{{Acharya} et~al.,}{{Acharya}
  et~al.}{2013}]{2013APh....43....3A}
{Acharya} B.~S.,  et~al., 2013, \mn@doi [Astroparticle Physics]
  {10.1016/j.astropartphys.2013.01.007}, \href
  {http://adsabs.harvard.edu/abs/2013APh....43....3A} {43, 3}

\bibitem[\protect\citeauthoryear{{Acharya} et~al.,}{{Acharya}
  et~al.}{2017}]{2017arXiv170907997C}
{Acharya} B.~S.,  et~al., 2017, preprint, \href
  {http://adsabs.harvard.edu/abs/2017arXiv170907997C} {} (\mn@eprint {arXiv}
  {1709.07997})

\bibitem[\protect\citeauthoryear{{Ackermann} et~al.,}{{Ackermann}
  et~al.}{2010}]{2010ApJ...716.1178A}
{Ackermann} M.,  et~al., 2010, \mn@doi [\apj] {10.1088/0004-637X/716/2/1178},
  \href {http://adsabs.harvard.edu/abs/2010ApJ...716.1178A} {716, 1178}

\bibitem[\protect\citeauthoryear{{Ackermann} et~al.,}{{Ackermann}
  et~al.}{2013}]{2013ApJS..209...11A}
{Ackermann} M.,  et~al., 2013, \mn@doi [\apjs] {10.1088/0067-0049/209/1/11},
  \href {http://adsabs.harvard.edu/abs/2013ApJS..209...11A} {209, 11}

\bibitem[\protect\citeauthoryear{{Ackermann} et~al.,}{{Ackermann}
  et~al.}{2014}]{2014Sci...343...42A}
{Ackermann} M.,  et~al., 2014, \mn@doi [Science] {10.1126/science.1242353},
  \href {http://adsabs.harvard.edu/abs/2014Sci...343...42A} {343, 42}

\bibitem[\protect\citeauthoryear{{Adri{\'a}n-Mart{\'{\i}}nez}
  et~al.,}{{Adri{\'a}n-Mart{\'{\i}}nez} et~al.}{2016}]{2016PhRvD..93l2010A}
{Adri{\'a}n-Mart{\'{\i}}nez} S.,  et~al., 2016, \mn@doi [\prd]
  {10.1103/PhysRevD.93.122010}, \href
  {http://adsabs.harvard.edu/abs/2016PhRvD..93l2010A} {93, 122010}

\bibitem[\protect\citeauthoryear{{Albert} et~al.,}{{Albert}
  et~al.}{2007}]{2007ApJ...667..358A}
{Albert} J.,  et~al., 2007, \mn@doi [\apj] {10.1086/520761}, \href
  {http://adsabs.harvard.edu/abs/2007ApJ...667..358A} {667, 358}

\bibitem[\protect\citeauthoryear{{Albert} et~al.,}{{Albert}
  et~al.}{2017a}]{2017PhRvD..96b2005A}
{Albert} A.,  et~al., 2017a, \mn@doi [\prd] {10.1103/PhysRevD.96.022005}, \href
  {http://adsabs.harvard.edu/abs/2017PhRvD..96b2005A} {96, 022005}

\bibitem[\protect\citeauthoryear{{Albert} et~al.,}{{Albert}
  et~al.}{2017b}]{2017ApJ...850L..35A}
{Albert} A.,  et~al., 2017b, \mn@doi [\apjl] {10.3847/2041-8213/aa9aed}, \href
  {http://adsabs.harvard.edu/abs/2017ApJ...850L..35A} {850, L35}

\bibitem[\protect\citeauthoryear{{Aleksi\'c} et~al.,}{{Aleksi\'c}
  et~al.}{2016}]{2016APh....43...76A}
{Aleksi\'c} J.,  et~al., 2016, \mn@doi [Astroparticle Physics]
  {10.1016/j.astropartphys.2015.02.005}, \href
  {http://adsabs.harvard.edu/abs/2016APh....72...76A} {72, 76}

\bibitem[\protect\citeauthoryear{{Ando} et~al.,}{{Ando}
  et~al.}{2013}]{2013RvMP...85.1401A}
{Ando} S.,  et~al., 2013, \mn@doi [Reviews of Modern Physics]
  {10.1103/RevModPhys.85.1401}, \href
  {http://adsabs.harvard.edu/abs/2013RvMP...85.1401A} {85, 1401}

\bibitem[\protect\citeauthoryear{Archambault et~al.,}{Archambault
  et~al.}{2017}]{Archambault201734}
Archambault S.,  et~al., 2017, \mn@doi [Astroparticle Physics]
  {http://doi.org/10.1016/j.astropartphys.2017.03.001}, 91, 34

\bibitem[\protect\citeauthoryear{Aso, Michimura, Somiya, Ando, Miyakawa,
  Sekiguchi, Tatsumi  \& Yamamoto}{Aso et~al.}{2013}]{PhysRevD.88.043007}
Aso Y.,  Michimura Y.,  Somiya K.,  Ando M.,  Miyakawa O.,  Sekiguchi T.,
  Tatsumi D.,   Yamamoto H.,  2013, \mn@doi [Phys. Rev. D]
  {10.1103/PhysRevD.88.043007}, 88, 043007

\bibitem[\protect\citeauthoryear{{Atwood} et~al.,}{{Atwood}
  et~al.}{2013}]{2013ApJ...774..76A}
{Atwood} W.~B.,  et~al., 2013, \mn@doi [\apj] {10.1088/0004-637X/774/1/76},
  \href {http://adsabs.harvard.edu/abs/2013ApJ...774..76A} {774, 76}

\bibitem[\protect\citeauthoryear{{Band} et~al.,}{{Band}
  et~al.}{1993}]{1993ApJ...413..281B}
{Band} B.,  et~al., 1993, \mn@doi [\apj] {10.1086/172995}, \href
  {http://adsabs.harvard.edu/abs/1993ApJ...413..281B} {413, 281}

\bibitem[\protect\citeauthoryear{{Bartos} \& {M{\'a}rka}}{{Bartos} \&
  {M{\'a}rka}}{2015}]{2015arXiv150900983B}
{Bartos} I.,  {M{\'a}rka} S.,  2015, preprint, \href
  {http://adsabs.harvard.edu/abs/2015arXiv150900983B} {} (\mn@eprint {arXiv}
  {1509.00983})

\bibitem[\protect\citeauthoryear{{Bartos}, {Brady}  \& {M{\'a}rka}}{{Bartos}
  et~al.}{2013}]{2013CQGra..30l3001B}
{Bartos} I.,  {Brady} P.,   {M{\'a}rka} S.,  2013, \mn@doi [Classical and
  Quantum Gravity] {10.1088/0264-9381/30/12/123001}, \href
  {http://adsabs.harvard.edu/abs/2013CQGra..30l3001B} {30, 123001}

\bibitem[\protect\citeauthoryear{{Bartos} et~al.,}{{Bartos}
  et~al.}{2014}]{2014MNRAS.443..738B}
{Bartos} I.,  et~al., 2014, \mn@doi [\mnras] {10.1093/mnras/stu1205}, \href
  {http://adsabs.harvard.edu/abs/2014MNRAS.443..738B} {443, 738}

\bibitem[\protect\citeauthoryear{{Bartos}, {Haiman}, {Marka}, {Metzger},
  {Stone}  \& {Marka}}{{Bartos} et~al.}{2017a}]{2017arXiv170102328B}
{Bartos} I.,  {Haiman} Z.,  {Marka} Z.,  {Metzger} B.~D.,  {Stone} N.~C.,
  {Marka} S.,  2017a, Nature Commun., \href
  {http://adsabs.harvard.edu/abs/2017arXiv170102328B} {8, 831}

\bibitem[\protect\citeauthoryear{{Bartos}, {Kocsis}, {Haiman}  \&
  {M{\'a}rka}}{{Bartos} et~al.}{2017b}]{2017ApJ...835..165B}
{Bartos} I.,  {Kocsis} B.,  {Haiman} Z.,   {M{\'a}rka} S.,  2017b, \mn@doi
  [\apj] {10.3847/1538-4357/835/2/165}, \href
  {http://adsabs.harvard.edu/abs/2017ApJ...835..165B} {835, 165}

\bibitem[\protect\citeauthoryear{{Beloborodov}, {Hasco{\"e}t}  \&
  {Vurm}}{{Beloborodov} et~al.}{2014}]{2014ApJ...788...36B}
{Beloborodov} A.~M.,  {Hasco{\"e}t} R.,   {Vurm} I.,  2014, \mn@doi [\apj]
  {10.1088/0004-637X/788/1/36}, \href
  {http://adsabs.harvard.edu/abs/2014ApJ...788...36B} {788, 36}

\bibitem[\protect\citeauthoryear{{Berger}}{{Berger}}{2014}]{2014ARA&A..52...43B}
{Berger} E.,  2014, \mn@doi [\araa] {10.1146/annurev-astro-081913-035926},
  \href {http://adsabs.harvard.edu/abs/2014ARA%26A..52...43B} {52, 43}

\bibitem[\protect\citeauthoryear{{Bernl{\"o}hr} et~al.,}{{Bernl{\"o}hr}
  et~al.}{2013}]{2013APh....43..171B}
{Bernl{\"o}hr} K.,  et~al., 2013, \mn@doi [Astroparticle Physics]
  {10.1016/j.astropartphys.2012.10.002}, \href
  {http://adsabs.harvard.edu/abs/2013APh....43..171B} {43, 171}

\bibitem[\protect\citeauthoryear{{Biland} et~al.,}{{Biland}
  et~al.}{2014}]{2014JInst...9P0012B}
{Biland} A.,  et~al., 2014, \mn@doi [Journal of Instrumentation]
  {10.1088/1748-0221/9/10/P10012}, \href
  {http://adsabs.harvard.edu/abs/2014JInst...9P0012B} {9, P10012}

\bibitem[\protect\citeauthoryear{{Blanchet}}{{Blanchet}}{2014}]{2014LRR....17....2B}
{Blanchet} L.,  2014, \mn@doi [Living Reviews in Relativity]
  {10.12942/lrr-2014-2}, \href
  {http://adsabs.harvard.edu/abs/2014LRR....17....2B} {17, 2}

\bibitem[\protect\citeauthoryear{{Boh{\'e}} et~al.,}{{Boh{\'e}}
  et~al.}{2017}]{2017PhRvD..95d4028B}
{Boh{\'e}} A.,  et~al., 2017, \mn@doi [\prd] {10.1103/PhysRevD.95.044028},
  \href {http://adsabs.harvard.edu/abs/2017PhRvD..95d4028B} {95, 044028}

\bibitem[\protect\citeauthoryear{Carosi, Ansoldi, Antonelli, Berti, De~Lotto,
  Longo  \& Stamerra}{Carosi et~al.}{2017}]{Carosi:2017tbk}
Carosi A.,  Ansoldi S.,  Antonelli L.~A.,  Berti A.,  De~Lotto B.,  Longo F.,
  Stamerra A.,  2017, \mn@doi [AIP Conf. Proc.] {10.1063/1.4968997}, 1792,
  060014

\bibitem[\protect\citeauthoryear{{Chan}, {Hu}, {Messenger}, {Hendry}  \&
  {Heng}}{{Chan} et~al.}{2017}]{2017ApJ...834...84C}
{Chan} M.~L.,  {Hu} Y.-M.,  {Messenger} C.,  {Hendry} M.,   {Heng} I.~S.,
  2017, \mn@doi [\apj] {10.3847/1538-4357/834/1/84}, \href
  {http://adsabs.harvard.edu/abs/2017ApJ...834...84C} {834, 84}

\bibitem[\protect\citeauthoryear{{Ciolfi} \& {Siegel}}{{Ciolfi} \&
  {Siegel}}{2015}]{2015arXiv150501420C}
{Ciolfi} R.,  {Siegel} D.~M.,  2015, preprint, \href
  {http://adsabs.harvard.edu/abs/2015arXiv150501420C} {} (\mn@eprint {arXiv}
  {1505.01420})

\bibitem[\protect\citeauthoryear{{Connaughton} et~al.,}{{Connaughton}
  et~al.}{2016}]{2016ApJ...826L..6C}
{Connaughton} V.,  et~al., 2016, \mn@doi [\apjl] {10.3847/2041-8205/826/1/L6},
  \href {http://adsabs.harvard.edu/abs/2016ApJ...826L..6C} {826, L6}

\bibitem[\protect\citeauthoryear{{Coughlin} \& {Stubbs}}{{Coughlin} \&
  {Stubbs}}{2016}]{2016ExA....42..165C}
{Coughlin} M.,  {Stubbs} C.,  2016, \mn@doi [Experimental Astronomy]
  {10.1007/s10686-016-9503-4}, \href
  {http://adsabs.harvard.edu/abs/2016ExA....42..165C} {42, 165}

\bibitem[\protect\citeauthoryear{{Coulter} et~al.,}{{Coulter}
  et~al.}{2017}]{2017arXiv171005452C}
{Coulter} D.~A.,  et~al., 2017, preprint, \href
  {http://adsabs.harvard.edu/abs/2017arXiv171005452C} {} (\mn@eprint {arXiv}
  {1710.05452})

\bibitem[\protect\citeauthoryear{{De Pasquale} et~al.,}{{De Pasquale}
  et~al.}{2010}]{2010ApJ...709L.146D}
{De Pasquale} M.,  et~al., 2010, \mn@doi [\apjl]
  {10.1088/2041-8205/709/2/L146}, \href
  {http://adsabs.harvard.edu/abs/2010ApJ...709L.146D} {709, L146}

\bibitem[\protect\citeauthoryear{{Dom{\'{\i}}nguez} et~al.,}{{Dom{\'{\i}}nguez}
  et~al.}{2011}]{2011MNRAS.410.2556D}
{Dom{\'{\i}}nguez} A.,  et~al., 2011, \mn@doi [\mnras]
  {10.1111/j.1365-2966.2010.17631.x}, \href
  {http://adsabs.harvard.edu/abs/2011MNRAS.410.2556D} {410, 2556}

\bibitem[\protect\citeauthoryear{{Fioretti} et~al.,}{{Fioretti}
  et~al.}{2015}]{2015arXiv150901943F}
{Fioretti} V.,  et~al., 2015, in Proceedings of the 34th International Cosmic
  Ray Conference: The Hague, The Netherlands, July 30-August 6, 2015.
  (arXiv:1509.01943)

\bibitem[\protect\citeauthoryear{{Fioretti}, {Bulgarelli}, {Sch\"ussler}
  et~al.}{{Fioretti} et~al.}{2016}]{2016SPIE.9906E.3OF}
{Fioretti} V.,  {Bulgarelli} A.,  {Sch\"ussler} F.,   et~al., 2016, in
  Proceedings of the Conference "Ground-based and Airborne Telescopes VI".
  Proceedings of the SPIE (arXiv:1608.04992)

\bibitem[\protect\citeauthoryear{{Funk}, {Hinton}  \& {CTA Consortium}}{{Funk}
  et~al.}{2013}]{2013APh....43..348F}
{Funk} S.,  {Hinton} J.~A.,   {CTA Consortium} 2013, \mn@doi [Astroparticle
  Physics] {10.1016/j.astropartphys.2012.05.018}, \href
  {http://adsabs.harvard.edu/abs/2013APh....43..348F} {43, 348}

\bibitem[\protect\citeauthoryear{{Gehrels} \& {M{\'e}sz{\'a}ros}}{{Gehrels} \&
  {M{\'e}sz{\'a}ros}}{2012}]{2012Sci...337..932G}
{Gehrels} N.,  {M{\'e}sz{\'a}ros} P.,  2012, \mn@doi [Science]
  {10.1126/science.1216793}, \href
  {http://adsabs.harvard.edu/abs/2012Sci...337..932G} {337, 932}

\bibitem[\protect\citeauthoryear{{Gehrels}, {Cannizzo}, {Kanner}, {Kasliwal},
  {Nissanke}  \& {Singer}}{{Gehrels} et~al.}{2016}]{2016ApJ...820..136G}
{Gehrels} N.,  {Cannizzo} J.~K.,  {Kanner} J.,  {Kasliwal} M.~M.,  {Nissanke}
  S.,   {Singer} L.~P.,  2016, \mn@doi [\apj] {10.3847/0004-637X/820/2/136},
  \href {http://adsabs.harvard.edu/abs/2016ApJ...820..136G} {820, 136}

\bibitem[\protect\citeauthoryear{{Ghosh}, {Bloemen}, {Nelemans}, {Groot}  \&
  {Price}}{{Ghosh} et~al.}{2016}]{2016A&A...592A..82G}
{Ghosh} S.,  {Bloemen} S.,  {Nelemans} G.,  {Groot} P.~J.,   {Price} L.~R.,
  2016, \mn@doi [\aap] {10.1051/0004-6361/201527712}, \href
  {http://adsabs.harvard.edu/abs/2016A%26A...592A..82G} {592, A82}

\bibitem[\protect\citeauthoryear{{Giuliani} et~al.,}{{Giuliani}
  et~al.}{2010}]{2010ApJ...708L..84G}
{Giuliani} A.,  et~al., 2010, \mn@doi [\apjl] {10.1088/2041-8205/708/2/L84},
  \href {http://adsabs.harvard.edu/abs/2010ApJ...708L..84G} {708, L84}

\bibitem[\protect\citeauthoryear{{Goldstein} et~al.,}{{Goldstein}
  et~al.}{2017}]{2017ApJ...848L..14G}
{Goldstein} A.,  et~al., 2017, \mn@doi [\apjl] {10.3847/2041-8213/aa8f41},
  \href {http://adsabs.harvard.edu/abs/2017ApJ...848L..14G} {848, L14}

\bibitem[\protect\citeauthoryear{{Greiner}, {Burgess}, {Savchenko}  \&
  {Yu}}{{Greiner} et~al.}{2016}]{2016ApJ...827L..38G}
{Greiner} J.,  {Burgess} J.~M.,  {Savchenko} V.,   {Yu} H.~F.,  2016, \mn@doi
  [\apjl] {10.3847/2041-8205/827/2/L38}, \href
  {http://adsabs.harvard.edu/abs/2016ApJ...827L..38G} {827, L38}

\bibitem[\protect\citeauthoryear{{Hasco\"et}, {Vurm}  \&
  {Beloborodov}}{{Hasco\"et} et~al.}{2015}]{2015ApJ...813...63H}
{Hasco\"et} R.,  {Vurm} I.,   {Beloborodov} A.~M.,  2015, \mn@doi [\apj]
  {10.1088/0004-637X/813/1/63}, \href
  {http://adsabs.harvard.edu/abs/2015ApJ...813...63H} {813, 63}

\bibitem[\protect\citeauthoryear{Hassan, Humensky, Nieto, Wood  et~al.}{Hassan
  et~al.}{2015}]{Hassan:2015yva}
Hassan T.,  Humensky B.,  Nieto D.,  Wood M.,   et~al., 2015, in {Proceedings
  of the 34th International Cosmic Ray Conference: The Hague, The Netherlands,
  July 30-August 6, 2015}. (arXiv:1508.06076)

\bibitem[\protect\citeauthoryear{{Hauser} \& {Dwek}}{{Hauser} \&
  {Dwek}}{2001}]{2001ARA&A..39..249H}
{Hauser} M.~G.,  {Dwek} E.,  2001, \mn@doi [\araa]
  {10.1146/annurev.astro.39.1.249}, \href
  {http://adsabs.harvard.edu/abs/2001ARA%26A..39..249H} {39, 249}

\bibitem[\protect\citeauthoryear{{IceCube Collaboration}}{{IceCube
  Collaboration}}{2013}]{2013Sci...342E...1I}
{IceCube Collaboration} 2013, \mn@doi [Science] {10.1126/science.1242856},
  \href {http://adsabs.harvard.edu/abs/2013Sci...342E...1I} {342, 1242856}

\bibitem[\protect\citeauthoryear{{Inoue} et~al.,}{{Inoue}
  et~al.}{2013a}]{2013APh....43..252I}
{Inoue} S.,  et~al., 2013a, \mn@doi [Astroparticle Physics]
  {10.1016/j.astropartphys.2013.01.004}, \href
  {http://adsabs.harvard.edu/abs/2013APh....43..252I} {43, 252}

\bibitem[\protect\citeauthoryear{{Inoue}, {Inoue}, {Kobayashi}, {Makiya},
  {Niino}  \& {Totani}}{{Inoue} et~al.}{2013b}]{2013ApJ...768..197I}
{Inoue} Y.,  {Inoue} S.,  {Kobayashi} M.~A.~R.,  {Makiya} R.,  {Niino} Y.,
  {Totani} T.,  2013b, \mn@doi [\apj] {10.1088/0004-637X/768/2/197}, \href
  {http://adsabs.harvard.edu/abs/2013ApJ...768..197I} {768, 197}

\bibitem[\protect\citeauthoryear{Iyer et~al.}{Iyer et~al.}{2011}]{LIGOIndia}
Iyer B.,  et~al., 2011, Technical report, LIGO-India

\bibitem[\protect\citeauthoryear{{Janiuk}, {Charzy\'nski}  \&
  {Bejger}}{{Janiuk} et~al.}{2013}]{2013A&A...560A..25J}
{Janiuk} A.,  {Charzy\'nski} S.,   {Bejger} M.,  2013, \mn@doi [\aap]
  {10.1051/0004-6361/201322165}, \href
  {http://adsabs.harvard.edu/abs/2013A%26A...560A..25J} {560, A25}

\bibitem[\protect\citeauthoryear{{Kanner}, {Huard}, {M{\'a}rka}, {Murphy},
  {Piscionere}, {Reed}  \& {Shawhan}}{{Kanner}
  et~al.}{2008}]{2008CQGra..25r4034K}
{Kanner} J.,  {Huard} T.~L.,  {M{\'a}rka} S.,  {Murphy} D.~C.,  {Piscionere}
  J.,  {Reed} M.,   {Shawhan} P.,  2008, \mn@doi [Classical and Quantum
  Gravity] {10.1088/0264-9381/25/18/184034}, \href
  {http://adsabs.harvard.edu/abs/2008CQGra..25r4034K} {25, 184034}

\bibitem[\protect\citeauthoryear{{Khan}, {Husa}, {Hannam}, {Ohme},
  {P{\"u}rrer}, {Forteza}  \& {Boh{\'e}}}{{Khan}
  et~al.}{2016}]{2016PhRvD..93d4007K}
{Khan} S.,  {Husa} S.,  {Hannam} M.,  {Ohme} F.,  {P{\"u}rrer} M.,  {Forteza}
  X.~J.,   {Boh{\'e}} A.,  2016, \mn@doi [\prd] {10.1103/PhysRevD.93.044007},
  \href {http://adsabs.harvard.edu/abs/2016PhRvD..93d4007K} {93, 044007}

\bibitem[\protect\citeauthoryear{{Kobayashi} \& {M\'esz\'aros}}{{Kobayashi} \&
  {M\'esz\'aros}}{2003}]{2003ApJ...589..861K}
{Kobayashi} S.,  {M\'esz\'aros} P.,  2003, \mn@doi [\apj] {10.1086/374733},
  \href {http://adsabs.harvard.edu/abs/2003ApJ...589..861K} {589, 861}

\bibitem[\protect\citeauthoryear{{LIGO Scientific Collaboration}}{{LIGO
  Scientific Collaboration}}{2015}]{LIGOPLUS}
{LIGO Scientific Collaboration} 2015, Technical report, LIGO-T1500290

\bibitem[\protect\citeauthoryear{{LIGO Scientific Collaboration} et~al.,}{{LIGO
  Scientific Collaboration} et~al.}{2012}]{2012A&A...539A.124L}
{LIGO Scientific Collaboration} et~al., 2012, \mn@doi [\aap]
  {10.1051/0004-6361/201118219}, \href
  {http://adsabs.harvard.edu/abs/2012A%26A...539A.124L} {539, A124}

\bibitem[\protect\citeauthoryear{{Lazzati}, {Perna}, {Morsony},
  {L\'opez-C\'amara}, {Cantiello}, {Ciolfi}, {Giacomazzo}  \&
  {Workman}}{{Lazzati} et~al.}{2017}]{2017arXiv171203237L}
{Lazzati} D.,  {Perna} R.,  {Morsony} B.~J.,  {L\'opez-C\'amara} D.,
  {Cantiello} M.,  {Ciolfi} R.,  {Giacomazzo} B.,   {Workman} J.~C.,  2017,
  preprint, \href {http://adsabs.harvard.edu/abs/2017arXiv171203237L} {}
  (\mn@eprint {arXiv} {1712.03237})

\bibitem[\protect\citeauthoryear{{Letessier-Selvon} \&
  {Stanev}}{{Letessier-Selvon} \& {Stanev}}{2011}]{2011RvMP...83..907L}
{Letessier-Selvon} A.,  {Stanev} T.,  2011, \mn@doi [Reviews of Modern Physics]
  {10.1103/RevModPhys.83.907}, \href
  {http://adsabs.harvard.edu/abs/2011RvMP...83..907L} {83, 907}

\bibitem[\protect\citeauthoryear{{Loeb}}{{Loeb}}{2016}]{2016ApJ...819L..21L}
{Loeb} A.,  2016, \mn@doi [\apjl] {10.3847/2041-8205/819/2/L21}, \href
  {http://adsabs.harvard.edu/abs/2016ApJ...819L..21L} {819, L21}

\bibitem[\protect\citeauthoryear{{Lyutikov}}{{Lyutikov}}{2016}]{2016arXiv160207352L}
{Lyutikov} M.,  2016, preprint, \href
  {http://adsabs.harvard.edu/abs/2016arXiv160207352L} {} (\mn@eprint {arXiv}
  {1602.07352})

\bibitem[\protect\citeauthoryear{{Maier}, {Arrabito}, {Bernl{\"o}hr},
  {Bregeon}, {Cumani}, {Hassan}, {Moralejo}  et~al.}{{Maier}
  et~al.}{2017}]{2017arXiv170901381M}
{Maier} G.,  {Arrabito} L.,  {Bernl{\"o}hr} K.,  {Bregeon} J.,  {Cumani} P.,
  {Hassan} T.,  {Moralejo} A.,   et~al., 2017, in Proceedings of the 35th
  International Cosmic Ray Conference: Busan, South Korea, July 12-20, 2017.
  (arXiv:1709.01381)

\bibitem[\protect\citeauthoryear{{Margalit}, {Metzger}  \&
  {Beloborodov}}{{Margalit} et~al.}{2015}]{2015PhRvL.115q1101M}
{Margalit} B.,  {Metzger} B.~D.,   {Beloborodov} A.~M.,  2015, \mn@doi
  [Physical Review Letters] {10.1103/PhysRevLett.115.171101}, \href
  {http://adsabs.harvard.edu/abs/2015PhRvL.115q1101M} {115, 171101}

\bibitem[\protect\citeauthoryear{{Meszaros} \& {Rees}}{{Meszaros} \&
  {Rees}}{1994}]{1994MNRAS.269L..41M}
{Meszaros} P.,  {Rees} M.~J.,  1994, \mn@doi [\mnras]
  {10.1093/mnras/269.1.L41}, \href
  {http://adsabs.harvard.edu/abs/1994MNRAS.269L..41M} {269, L41}

\bibitem[\protect\citeauthoryear{{Meszaros}, {Rees}  \&
  {Papathanassiou}}{{Meszaros} et~al.}{1994}]{1994ApJ...432..181M}
{Meszaros} P.,  {Rees} M.~J.,   {Papathanassiou} H.,  1994, \mn@doi [\apj]
  {10.1086/174559}, \href {http://adsabs.harvard.edu/abs/1994ApJ...432..181M}
  {432, 181}

\bibitem[\protect\citeauthoryear{{Montaruli}, {Greenshaw}, {Sol}, {Pareschi}
  et~al.}{{Montaruli} et~al.}{2015}]{Montaruli:2015xya}
{Montaruli} T.,  {Greenshaw} T.,  {Sol} H.,  {Pareschi} G.,   et~al., 2015, in
  Proceedings of the 34th International Cosmic Ray Conference: The Hague, The
  Netherlands, July 30-August 6, 2015. (arXiv:1508.06472)

\bibitem[\protect\citeauthoryear{{Murase}, {Kashiyama}, {M{\'e}sz{\'a}ros},
  {Shoemaker}  \& {Senno}}{{Murase} et~al.}{2016}]{2016ApJ...822L...9M}
{Murase} K.,  {Kashiyama} K.,  {M{\'e}sz{\'a}ros} P.,  {Shoemaker} I.,
  {Senno} N.,  2016, \mn@doi [\apjl] {10.3847/2041-8205/822/1/L9}, \href
  {http://adsabs.harvard.edu/abs/2016ApJ...822L...9M} {822, L9}

\bibitem[\protect\citeauthoryear{{Ong}}{{Ong}}{1998}]{1998PhR...305...93O}
{Ong} R.~A.,  1998, \mn@doi [\physrep] {10.1016/S0370-1573(98)00026-X}, \href
  {http://adsabs.harvard.edu/abs/1998PhR...305...93O} {305, 93}

\bibitem[\protect\citeauthoryear{Otte et~al.}{Otte et~al.}{2015}]{Otte:2015mya}
Otte A.~N.,  et~al., 2015, in {Proceedings, 34th International Cosmic Ray
  Conference (ICRC 2015): The Hague, The Netherlands, July 30-August 6, 2015}.
  (arXiv:1509.02345)

\bibitem[\protect\citeauthoryear{{Perna}, {Lazzati}  \& {Giacomazzo}}{{Perna}
  et~al.}{2016}]{2016ApJ...821L..18P}
{Perna} R.,  {Lazzati} D.,   {Giacomazzo} B.,  2016, \mn@doi [\apjl]
  {10.3847/2041-8205/821/1/L18}, \href
  {http://adsabs.harvard.edu/abs/2016ApJ...821L..18P} {821, L18}

\bibitem[\protect\citeauthoryear{{Piran}}{{Piran}}{2004}]{2004RvMP...76.1143P}
{Piran} T.,  2004, \mn@doi [Reviews of Modern Physics]
  {10.1103/RevModPhys.76.1143}, \href
  {http://adsabs.harvard.edu/abs/2004RvMP...76.1143P} {76, 1143}

\bibitem[\protect\citeauthoryear{{Piscionere}, {Marka}, {Shawhan}, {Kanner},
  {Huard}  \& {Murphy}}{{Piscionere} et~al.}{2007}]{2007AAS...211.9903P}
{Piscionere} J.,  {Marka} S.,  {Shawhan} P.~S.,  {Kanner} J.,  {Huard} T.~L.,
  {Murphy} D.~C.,  2007, in American Astronomical Society Meeting Abstracts.
  p.~910

\bibitem[\protect\citeauthoryear{{Pretorius}}{{Pretorius}}{2005}]{2005PhRvL..95l1101P}
{Pretorius} F.,  2005, \mn@doi [Physical Review Letters]
  {10.1103/PhysRevLett.95.121101}, \href
  {http://adsabs.harvard.edu/abs/2005PhRvL..95l1101P} {95, 121101}

\bibitem[\protect\citeauthoryear{{Salafia}, {Colpi}, {Branchesi},
  {Chassande-Mottin}, {Ghirlanda}, {Ghisellini}  \& {Vergani}}{{Salafia}
  et~al.}{2017}]{2017ApJ...846...62S}
{Salafia} O.~S.,  {Colpi} M.,  {Branchesi} M.,  {Chassande-Mottin} E.,
  {Ghirlanda} G.,  {Ghisellini} G.,   {Vergani} S.~D.,  2017, \mn@doi [\apj]
  {10.3847/1538-4357/aa850e}, \href
  {http://adsabs.harvard.edu/abs/2017ApJ...846...62S} {846, 62}

\bibitem[\protect\citeauthoryear{Santander}{Santander}{2016}]{Santander:2016chw}
Santander M.,  2016, in {Proceedings, 38th International Conference on High
  Energy Physics (ICHEP 2016): Chicago, IL, USA, August 3-10, 2016}. SISSA
  (\mn@eprint {arXiv} {1612.04301}), \url
  {https://inspirehep.net/record/1503147/files/arXiv:1612.04301.pdf}

\bibitem[\protect\citeauthoryear{Sathyaprakash \& Schutz}{Sathyaprakash \&
  Schutz}{2009}]{Sathyaprakash2009}
Sathyaprakash B.~S.,  Schutz B.~F.,  2009, \mn@doi [Living Reviews in
  Relativity] {10.12942/lrr-2009-2}, 12, 2

\bibitem[\protect\citeauthoryear{{Savchenko} et~al.,}{{Savchenko}
  et~al.}{2016}]{2016ApJ...820L..36S}
{Savchenko} V.,  et~al., 2016, \mn@doi [\apjl] {10.3847/2041-8205/820/2/L36},
  \href {http://adsabs.harvard.edu/abs/2016ApJ...820L..36S} {820, L36}

\bibitem[\protect\citeauthoryear{{Savchenko} et~al.,}{{Savchenko}
  et~al.}{2017}]{2017ApJ...848L..15S}
{Savchenko} V.,  et~al., 2017, \mn@doi [\apjl] {10.3847/2041-8213/aa8f94},
  \href {http://adsabs.harvard.edu/abs/2017ApJ...848L..15S} {848, L15}

\bibitem[\protect\citeauthoryear{{Seglar-Arroyo}, {Sch{\"u}ssler }
  et~al.}{{Seglar-Arroyo} et~al.}{2017}]{2017arXiv170510138S}
{Seglar-Arroyo} M.,  {Sch{\"u}ssler } F.,   et~al., 2017, preprint, \href
  {http://adsabs.harvard.edu/abs/2017arXiv170510138S} {} (\mn@eprint {arXiv}
  {1705.10138})

\bibitem[\protect\citeauthoryear{{Singer} et~al.,}{{Singer}
  et~al.}{2014}]{2014ApJ...795..105S}
{Singer} L.~P.,  et~al., 2014, \mn@doi [\apj] {10.1088/0004-637X/795/2/105},
  \href {http://adsabs.harvard.edu/abs/2014ApJ...795..105S} {795, 105}

\bibitem[\protect\citeauthoryear{{Singer} et~al.,}{{Singer}
  et~al.}{2016}]{2016ApJ...829L..15S}
{Singer} L.~P.,  et~al., 2016, \mn@doi [\apjl] {10.3847/2041-8205/829/1/L15},
  \href {http://adsabs.harvard.edu/abs/2016ApJ...829L..15S} {829, L15}

\bibitem[\protect\citeauthoryear{{Smith} et~al.,}{{Smith}
  et~al.}{2013}]{2013APh....45...56S}
{Smith} M.~W.~E.,  et~al., 2013, \mn@doi [Astroparticle Physics]
  {10.1016/j.astropartphys.2013.03.003}, \href
  {http://adsabs.harvard.edu/abs/2013APh....45...56S} {45, 56}

\bibitem[\protect\citeauthoryear{{Stone}, {Metzger}  \& {Haiman}}{{Stone}
  et~al.}{2016}]{2016arXiv160204226S}
{Stone} N.~C.,  {Metzger} B.~D.,   {Haiman} Z.,  2016, preprint, \href
  {http://adsabs.harvard.edu/abs/2016arXiv160204226S} {} (\mn@eprint {arXiv}
  {1602.04226})

\bibitem[\protect\citeauthoryear{{Szanecki}, {Sobczy{\'n}ska},
  {Nied{\'z}wiecki}, {Sitarek}  \& {Bednarek}}{{Szanecki}
  et~al.}{2015}]{2015APh....67...33S}
{Szanecki} M.,  {Sobczy{\'n}ska} D.,  {Nied{\'z}wiecki} A.,  {Sitarek} J.,
  {Bednarek} W.,  2015, \mn@doi [Astroparticle Physics]
  {10.1016/j.astropartphys.2015.01.008}, \href
  {http://adsabs.harvard.edu/abs/2015APh....67...33S} {67, 33}

\bibitem[\protect\citeauthoryear{{Tavani} et~al.,}{{Tavani}
  et~al.}{2016}]{2016ApJ...825L...4T}
{Tavani} M.,  et~al., 2016, \mn@doi [\apjl] {10.3847/2041-8205/825/1/L4}, \href
  {http://adsabs.harvard.edu/abs/2016ApJ...825L...4T} {825, L4}

\bibitem[\protect\citeauthoryear{{The LIGO Scientific Collaboration}
  et~al.,}{{The LIGO Scientific Collaboration}
  et~al.}{2016}]{2016arXiv160604856T}
{The LIGO Scientific Collaboration} et~al., 2016, preprint, \href
  {http://adsabs.harvard.edu/abs/2016arXiv160604856T} {} (\mn@eprint {arXiv}
  {1606.04856})

\bibitem[\protect\citeauthoryear{{The LIGO Scientific Collaboration}
  et~al.,}{{The LIGO Scientific Collaboration}
  et~al.}{2017}]{2017arXiv170404628T}
{The LIGO Scientific Collaboration} et~al., 2017, preprint, \href
  {http://adsabs.harvard.edu/abs/2017arXiv170404628T} {} (\mn@eprint {arXiv}
  {1704.04628})

\bibitem[\protect\citeauthoryear{{Verrecchia} et~al.,}{{Verrecchia}
  et~al.}{2017}]{2017ApJ...847L..20V}
{Verrecchia} F.,  et~al., 2017, \apjl, \href
  {http://adsabs.harvard.edu/abs/2017ApJ...847L..20V} {847, L20}

\bibitem[\protect\citeauthoryear{{Vietri} \& {Stella}}{{Vietri} \&
  {Stella}}{1998}]{1998ApJ...507L..45V}
{Vietri} M.,  {Stella} L.,  1998, \mn@doi [\apjl] {10.1086/311674}, \href
  {http://adsabs.harvard.edu/abs/1998ApJ...507L..45V} {507, L45}

\bibitem[\protect\citeauthoryear{{Vurm} \& {Beloborodov}}{{Vurm} \&
  {Beloborodov}}{2017}]{2017ApJ...846..152V}
{Vurm} I.,  {Beloborodov} A.~M.,  2017, \mn@doi [\apj]
  {10.3847/1538-4357/aa7ddb}, \href
  {http://adsabs.harvard.edu/abs/2017ApJ...846..152V} {846, 152}

\bibitem[\protect\citeauthoryear{{Waxman} \& {Bahcall}}{{Waxman} \&
  {Bahcall}}{1997}]{1997PhRvL..78.2292W}
{Waxman} E.,  {Bahcall} J.,  1997, \mn@doi [Physical Review Letters]
  {10.1103/PhysRevLett.78.2292}, \href
  {http://adsabs.harvard.edu/abs/1997PhRvL..78.2292W} {78, 2292}

\bibitem[\protect\citeauthoryear{{Woosley} \& {Bloom}}{{Woosley} \&
  {Bloom}}{2006}]{2006ARA&A..44..507W}
{Woosley} S.~E.,  {Bloom} J.~S.,  2006, \mn@doi [\araa]
  {10.1146/annurev.astro.43.072103.150558}, \href
  {http://adsabs.harvard.edu/abs/2006ARA%26A..44..507W} {44, 507}

\bibitem[\protect\citeauthoryear{{Xiong}}{{Xiong}}{2016}]{2016arXiv160505447X}
{Xiong} S.,  2016, preprint, \href
  {http://adsabs.harvard.edu/abs/2016arXiv160505447X} {} (\mn@eprint {arXiv}
  {1605.05447})

\bibitem[\protect\citeauthoryear{{Yamazaki}, {Asano}  \& {Ohira}}{{Yamazaki}
  et~al.}{2016}]{2016PTEP.2016e1E01Y}
{Yamazaki} R.,  {Asano} K.,   {Ohira} Y.,  2016, Progress of Theoretical and
  Experimental Physics, \href
  {http://adsabs.harvard.edu/abs/2016PTEP.2016e1E01Y} {2016, 051E01}

\bibitem[\protect\citeauthoryear{{Zhu} \& {Wang}}{{Zhu} \&
  {Wang}}{2016}]{2016ApJ...828L...4Z}
{Zhu} Q.-Y.,  {Wang} X.-Y.,  2016, \mn@doi [\apjl]
  {10.3847/2041-8205/828/1/L4}, \href
  {http://adsabs.harvard.edu/abs/2016ApJ...828L...4Z} {828, L4}

\bibitem[\protect\citeauthoryear{{van Putten}, {Levinson}, {Lee}, {Regimbau},
  {Punturo}  \& {Harry}}{{van Putten} et~al.}{2004}]{2004PhRvD..69d4007V}
{van Putten} M.~H.,  {Levinson} A.,  {Lee} H.~K.,  {Regimbau} T.,  {Punturo}
  M.,   {Harry} G.~M.,  2004, \mn@doi [\prd] {10.1103/PhysRevD.69.044007},
  \href {http://adsabs.harvard.edu/abs/2004PhRvD.69d4007V} {69, 044007}

\makeatother
\end{thebibliography}

\bsp	
\label{lastpage}
\end{document}